\documentclass[aps,prb,twocolumn,showpacs,groupedaddress]{revtex4-1}
\usepackage{graphicx}% Include figure files
\usepackage{dcolumn}% Align table columns on decimal point
\usepackage{bm}% bold math
\usepackage{amsmath}% Split the equation
\usepackage{natbib}
\usepackage{lineno}
 \usepackage{array}

\begin{document}

\preprint{APS/123-QED}

\title{Field-induced domain wall propagation: beyond the one-dimensional model}

\author{L. Thevenard$^{1}$\email[e-mail: ]{thevenard@insp.jussieu.fr},  C. Gourdon$^{1}$, S. Haghgoo$^{1}$, J-P. Adam$^{2}$, H. J. von Bardeleben$^{1}$,  A. Lema\^itre$^{3}$, W. Schoch$^{4}$,  and A. Thiaville$^{2}$,}
 %\altaffiliation[Also at ]{}%Lines break automatically or can be forced with \\

%\homepage[web site: ]{http://www.insp.upmc.fr}
 
\affiliation{
$^1$ Institut des Nanosciences de Paris, Universit\'{e} Pierre et Marie Curie, CNRS, UMR7588, 4 place Jussieu,75252 Paris, France\\
$^2$ Laboratoire de Physique des Solides, CNRS - Universit\'{e} Paris-Sud, b\^at. 510, 91405 Orsay, France \\
$^3$  Laboratoire de Photonique et Nanostructures, CNRS, UPR 20, Route de Nozay, Marcoussis, 91460, France\\
$^4$ Institut f$\ddot{u}$r Quantenmaterie, Universit$\ddot{a}$t Ulm, 89069 Ulm, Germany
}

\date{\today}% It is always \today, today,

\label{sec:Abstract}

\begin{abstract}
%\section{Abstract}

 We have investigated numerically  the field-driven propagation of perpendicularly magnetized ferromagnetic layers. It was then compared  to the historical one-dimensional domain wall (DW) propagation model widely used in spintronics studies of magnetic nanostructures. In the particular regime of layer thickness ($h$) of the order of the exchange length, anomalous velocity peaks  appear in the precessional regime, their shape and position shifting with $h$. This has also been observed experimentally. Analyses of the simulations show a distinct correlation between the curvature of the DW and the twist of the magnetization vector within it, and the velocity peak. Associating a phenomenological description of this twist with a four-coordinate DW propagation model, we reproduce very well these kinks and show that they result from the torque exerted by  the stray field created by the domains  on the twisted magnetization. The position of the peaks is well predicted from the DW's first flexural mode  frequency, and  depends strongly on the layer thickness. Comparison of the proposed model to  DW propagation data obtained on dilute semiconductor ferromagnets GaMnAs and GaMnAsP sheds light on the origin of the measured peaks.

% abstract max 100 mots pour APL
\end{abstract}

%Magnetic semiconductors, DW and domain structure, Dynamics of domain structures 

\maketitle

\section{INTRODUCTION}

The 1970's  saw the establishment of the main theories describing magnetic domain wall (DW) propagation, among which the so-called  one-dimensional (1D) model\cite{Schryer1974,Slonczewski1973,Hagedorn1974,Thiele1974,Malozemoff1972}. These studies fueled the intense efforts towards the building of magnetic bubble memories. They have recently come back into fashion with new schemes being proposed to use domains or domain walls as the building blocks of a non-volatile and down-scalable memory\cite{SolitonPatent,Diegel2009,parkin08}.  Described in different formalisms\cite{Schryer1974,Slonczewski1973,Hagedorn1974,Thiele1974,Malozemoff1972}, the 1D model assumes that the DW propagation can be fully described by two time-dependent variables: the position $q_{0}(t)$ and  azimuthal angle $\phi_{0}(t)$ of the magnetization unit vector $\vec{m}$. For a defect-free sample, the velocity versus field curve v(H) is then shown to consist of a high mobility regime where the configuration of $\vec{m}$ within the DW  is stationary (${\phi_{0}}(t)$ constant) up to the Walker field $H_{W}$, followed by  a linear, lower mobility  regime where $\vec{m}$  precesses around the applied field.

While these two regimes have indeed been evidenced experimentally\cite{Metaxas2007a,Dourlat2008,Beach2005}, one or two unexpected kinks have repeatedly been observed in a variety of configurations: in-plane magnetized Permalloy\cite{Moriya2010},   out-of-plane magnetized garnets\cite{Vella75} as well as out-of-plane magnetized  ferromagnetic semiconductor GaMnAs\cite{Dourlat2008}. It has also been observed in field-assisted current-induced  propagation\cite{Hayashi2006}. The present work aims to address this issue, all the more crucial as the 1D model is now routinely used as a guideline for a vast number of studies on current-induced DW propagation\cite{beach2007,Thiaville2007,Adam2009a}. Whereas differences with the 1D model in the stationary regime were attributed quite early on to the nucleation/annihilation of Bloch lines\cite{Slonczewski1974}, those in the precessional regimes have, to our knowledge, only been considered by the numerical simulations of  Patterson et. al\cite{Patterson1991} who did not isolate the physical mechanism(s) for these kinks. 

\begin{figure}
	\centering
		\includegraphics[width=0.45\textwidth]{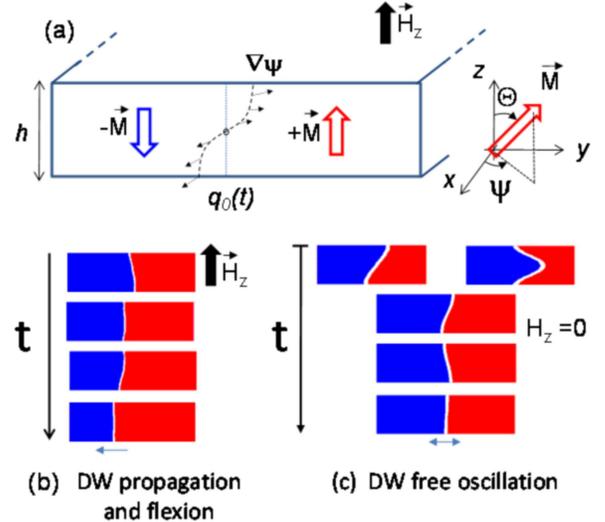}

	\caption{[color on-line] (a) Geometry of the simulation, side view of the 1$\mu$m$\times$$h$ layer. The DW propagates towards the left, its mean position along $y$ is given by $q_{0}$(t) (not to scale). (b) DW propagation evidencing curvature, shown here for $h$=80~nm to highlight the effect. (c) DW free oscillation ($H_{z}$=0), after launching from either of the top two  positions. }
	\label{fig:Fig1}
\end{figure}

In the following section of this paper (Section II), we  present  simulated v(H) curves for perpendicularly magnetized ferromagnetic layers of increasing thickness ($h$=10-40~nm) obtained with an open source micro-magnetic simulation software\cite{oommf}. In Sections III and IV, we analyze extensively the three component magnetic configuration within the DWs to relate it to the position of the v(H) peaks. In particular, we  evidence the role of the curvature of the DW and the twist of its magnetization vector  in increasing the DW velocity above the 1D model value for certain fields. Those particular  fields are found to correspond to the first flexural mode resonance of the  domain wall. Finally, in section V this  approach is tested against experimental v(H) curves obtained on GaMnAs and GaMnAsP thin films of varying thickness, magnetization, magnetic anisotropy and exchange constants. Ferromagnetic semiconductors are ideal for these studies as  their weak pinning allows all DW propagation regimes to be observed\cite{Dourlat2008}, their magnetic parameters can easily be tuned by careful doping\cite{Haghgoo2010,thevenard07,Cubukcu2010}, and the layer thickness can be varied across the exchange length. Section VI summarizes the results and opens new perspectives for DW propagation in other geometries.

\section{SIMULATIONS}

\subsection*{\label{sec:GeometryResults}A. Geometry \& results\protect\\ }

\begin{figure}
	\centering
			\includegraphics[width=0.40\textwidth]{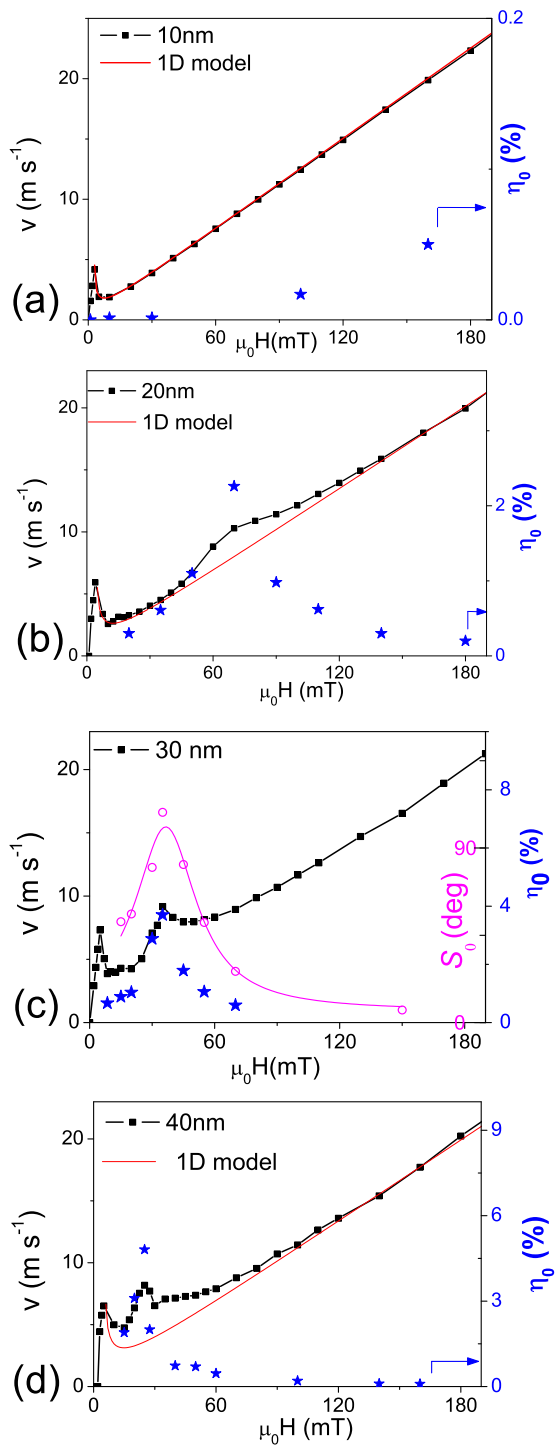}
	\caption{[color on-line] Simulated velocity versus field (squares): (a) $h$=10~nm, (b) $h$=20~nm, (c) $h$=30~nm, and (d) $h$=40~nm. In (a), (b) and (d), the solid line is a fit to the 1D model, and the stars  the maximum DW elongation $\eta_{0}$. In (c) the open circles are the twist amplitude $S_{0}$ and its Lorentzian fit (solid line). Investigated thicknesses lie on either side of the exchange length, $\pi \Lambda$=27~nm.}
	\label{fig:Fig2}
\end{figure}

The simulation was laid out on a 2D mesh with cell dimensions $c^{2}$,  and a strip of length\cite{striplength}  $d$=1$~\mu m$. Varying heights were investigated: $h$=10 nm ($c$=1~nm), and  $h$=20, 30, 40~nm ($c$=2~nm). The film was taken infinite in the $x$ direction (Fig. \ref{fig:Fig1}a). The applied field  $H$=$H_{z}$ was constant and  perpendicular to the plane of the layer, and  micromagnetic parameters typical of GaMnAs\cite{gourdon2007} were used: magnetization $M$=33~kA~m$^{-1}$, uniaxial anisotropy coefficient $K_{u}$=8878~J~m$^{-3}$ and field  $\mu_0 H_{u}=2K_{u}/M$, exchange constant $A$=5.10$^{-14}$~J~m$^{-1}$, and damping coefficient $\alpha$=0.3. Given the high value of the factor $Q=\frac{2K_{u}}{\mu_{0}M^{2}}$$>$3  in our layers, the complex in-plane anisotropy of GaMnAs was  ignored. Its influence will be  discussed later. The magnetization orientation vector $\vec{m}$ was  defined in Cartesian ($m_{x}$, $m_{y}$, $m_{z}$) or ($\Theta, \psi $) coordinates (Fig. \ref{fig:Fig1}a). The position coordinate $q_{0}$ is given as usual\cite{Malozemoff1972} by $\dot{\Theta}$=-$\frac{\dot{q_{0}}}{\Delta}\sin\Theta$ where the dot denotes the time derivative. The instantaneous velocity v$_{inst} =\frac{d}{2} \frac{\partial < m_{z}>}{\partial t}$ was averaged over many periods and the resulting v(H) curves are shown in Fig. \ref{fig:Fig2}. For comparison, the  baseline of the  curves was  fitted to the 1D model velocity (keeping $\alpha$ fixed to 0.3, and varying $\mu_{0}H_{W}$ and $\Delta$), using the  ($q_{0}$,$\phi_{0})$(t) solutions to the Landau-Lifshitz-Gilbert (LLG) equation\cite{Schryer1974} for $H>H_{W}$:

\begin{align}
  \dot{q_{0}}(t) & = \frac{ \gamma \Delta \mu_{0}M}{2(1+\alpha^{2})}\sin 2\phi_{0}+\frac{\alpha \gamma \Delta }{1+\alpha^{2}}\mu_0 H\\
 \phi_{0}(t)     & =\arctan \left[\frac{H_{W}}{H}+\sqrt{1-\left[\frac{H_{W}}{H}\right]^{2}}\tan(\frac{2\pi t}{T})\right]
 \label{eq:psi0}
\end{align}

In these expressions, $H_{W}=\alpha M/2$ is the Walker field for a film of infinite thickness, $\Delta$=$\sqrt{A/K}$  the static DW width and $T=\frac{1+\alpha^{2}}{\mu_{0} \gamma}\frac{2\pi}{\sqrt{H^{2}-H^{2}_{W}}}$  the precession period (time for $\phi_{0}(t)$ to span  360$^{\circ}$).

The simulations were done up to 300~mT, but only the first part of the curve is shown in Fig. \ref{fig:Fig2} as the velocity  maintains the same mobility for higher fields. For $h$=10~nm (Fig.\ref{fig:Fig2}a), the numerical simulations as well as the 1D model lead to a very standard curve. A high mobility is observed up to the Walker field, then a narrow negative mobility region followed by a positive lower mobility regime. The DW velocity reaches over 30~m~s$^{-1}$ at $\mu_{0} H$=250~mT. No kinks were observed in the 0-300~mT range. The $h$=20-40~nm curves evidence a similar behavior up to Walker breakdown, but distinctive kinks appear above: a broad peak ($\approx$70~mT) for $h$=20~nm, a slightly asymetric peak  ($\approx$37~mT) for $h$=30~nm, and  two close peaks (25 and 35~mT) for $h$=40~nm. Walker breakdown occurs at low field ($\mu_{0} H_{W}$=3~mT, velocity $V_{W}$=4~ms$^{-1}$) for the thinnest layer, and then gradually converges to the infinite thickness film value $\mu_{0} H_{W}$=6~mT for $h$=40~nm. In the mean-time,  $\Delta$ decreases slightly with thickness (from 2.6~nm to 2.3~nm), reflecting  the change of demagnetizing factors of the DW  with layer thickness, which in turn modifies the values of $H_{W}$ and $\Delta$\cite{mougin2007}. The precession frequency of the magnetization $f_{prec}$ obtained as the inverse of the period of $m_{x}$(t) was found identical for all thicknesses and linear with the applied field (solid line in Fig. \ref{fig:Fig5}).

\subsection*{\label{sec:ImageAnalysis}B. Image analysis\protect\\}

\begin{figure}
	\centering
				\includegraphics[width=0.4\textwidth]{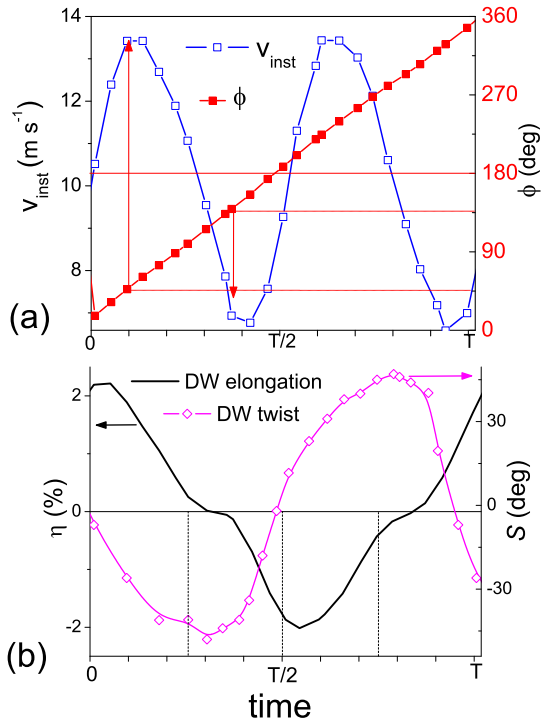}
	\caption{[color on-line] $h$=20~nm, $\mu_{0}H$=70~mT (a)  Instantaneous velocity (open squares) and mean $\phi$(t) (full squares) over one precession period. Arrows indicate maximum (minimum) velocities for the canting angle close to 45$^{\circ}$ [180$^{\circ}$]  (135$^{\circ}$ [180$^{\circ}$]). (b) Domain wall elongation (solid line) and domain wall twist $S$.}
	\label{fig:Fig3}
\end{figure}

The three component $\vec{m}$ vector field images (of the type in Fig. \ref{fig:Fig1}b,c) were then treated numerically to extract for each iteration the  domain wall width $\Delta$ and the depth-dependent DW position $q(z,t)$  and $\psi (z,t)$ angle  (averaged  over 2$\Delta$ along $y$). The  z-averaged angle $\phi$(t) is given by $\frac{1}{N_{i}}\sum^{N_{i}}_{1}\psi_{i} (z,t)$ with $N_{i}$  the number of cells across the thickness.  Plotting simultaneously  v$_{inst}$(t) and $\phi$(t) for $H>H_{W}$ over one precession period shows the main features of the precessional regime (Fig. \ref{fig:Fig3}a):   $\vec{m}$ precesses around the applied field inducing a demagnetizing field  across the wall. This sinusoidal field produces a torque on  $\vec{m}$ leading to an alternating forwards or backwards motion of the DW. The applied field term in Eq. (1) insures that the total wall displacement over one period is strictly positive. As expected from the 1D expression for $\phi_{0}(t)$ (Eqs. (1,2)), the instantaneous velocity is then maximum for  $\phi$=$45^{\circ}$~ modulo 180$^{\circ}$ and minimum for $\phi$=$135^{\circ}$~ [180$^{\circ}]$,  corresponding to the extrema of the demagnetizing torque (arrows in Fig. \ref{fig:Fig3}a). 

%The phase of this flexion varies slightly with respect to  the average angle $\phi$(t). 
Meanwhile, the DW length is not constant: the DW undergoes a $T$-periodic flexion (Fig. \ref{fig:Fig1}b and \ref{fig:Fig3}b). Let us define the length of the domain wall $l(t)$ as the curvilinear integral of $q(z,t)$ and its elongation as $\eta (t)=[l(t)-h]/h$. The simulations show that the DW goes from no elongation ($l$=$h$, $\eta$=0)  to the maximum  $\eta_{0}$ while $\phi$(t) varies from 0 to 360$^{\circ}$. Careful analysis shows that  the curvature is always of an $n$=1 mode type (following  the numbering of Slonczewski\cite{Slonczewski1981}), i.e odd about the layer's mid-height. For the thinnest $h$=10~nm layer, the amplitude of elongation increases progressively with field up to 1.53$\%$ for 300~mT (Fig. \ref{fig:Fig2}a).   For $h$=20-40~nm, $\eta_{0}$(H) reaches a maximum at a resonance field $H_{res}$ which decreases with layer thickness. This maximum elongation (stars in Fig. \ref{fig:Fig2}b,c,d) increases with $h$: $\eta_{0}$=2.3$\%$ for $h$=20~nm ($\mu_{0} H_{res}$=70~mT), 3.7$\%$ for $h$=30~nm ($\mu_{0} H_{res}$=35~mT) and $4.8\%$ for $h$=40~nm ($\mu_{0} H_{res}$=25~mT). Far from $H_{res}$, the flexion becomes negligible. Note that the $H_{res}$ fields correspond very exactly to those of the kinks in the v(H) curve. The enhancement of velocity therefore seems related to the field-dependent elongation of the DW.

\begin{figure}
	\centering
			\includegraphics[width=0.35\textwidth]{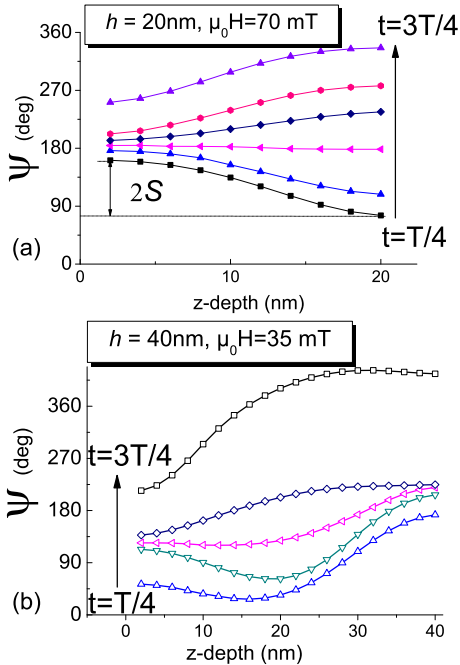}
	\caption{[color on-line] Depth dependence of $\psi$ (z,t) over half a precession period for (a) $h$=20~nm, $\mu_{0}H$=70~mT  and (b)  $h$=40~nm, $\mu_{0}H$=35~mT. Loose horizontal Bloch lines appear in the thicker layer.}
	\label{fig:Fig3bis}
\end{figure}

  For $h$=10-30~nm and $h$=40~nm when $\mu_{0}H<$30~mT, the twist amplitude is minimum  for $t$=T/4, maximum around $3T/4$, and null for $t$=0 [T/2] (Fig. \ref{fig:Fig3}b).  Twist and curvature are  intimately related, being in quadrature phase with respect to each other. The depth dependence $\psi (z,t)$ of the propagating DW evidences a clear twist from top to bottom of the layer (Fig. \ref{fig:Fig3bis}), but quite differently from the text-book  180$^{\circ}$ twist case of a static N\'eel-Bloch-N\'eel DW\cite{Hubert1975,Hagedorn1974}. Plotting $\psi(z,t)$ over a whole period shows that the twist is sinusoidal across the layer's thickness, and that its amplitude $S$(t) also varies sinusoidally in time over $T$ with a span of $S_{0}$ (Fig. \ref{fig:Fig3bis}a). The time and depth dependence of $\psi (z,t)$ are then described in  first approximation by the following  phenomenological expressions:

\begin{align}
     \psi(z,t) &=\phi_{0}(t)+\tilde{\psi}(z,t)\\     
    \tilde{\psi}(z,t) &  =   -S(t,H)\cos \frac{\pi z}{h}\\
     S(t,H)	&=-S_{0}(H)\sin \frac{2\pi t}{T(H)}  
 	\label{eq:psiz}
\end{align}

 The thickness dependence of $S_{0}$ is as expected: it increases with thickness, since  the exchange energy cost of a twist can be more easily accommodated in a thicker layer. Strikingly,  $S_{0}(H)$ follows the same trend as $\eta_{0}(H)$: it reaches a maximum at $H_{res}$ (see Fig. \ref{fig:Fig2}c for the $h$=30~nm case). At resonance, the full twist amplitude of the DW magnetization across the layer's thickness (2$S_{0}$) reaches about 90$^{\circ}$  for $h$=20~nm, 216$^{\circ}$ for $h$=30~nm and 230$^{\circ}$  for $h$=40~nm.

For the thickest layer $h$=40~nm,  a different behavior appears around $\mu_0 H$=30~mT, as can be seen in Fig. \ref{fig:Fig3bis}b: the twisting direction inverts  at a given depth. This corresponds to a horizontal Bloch line (HBL). These appear where the DW is most labile as a result  of the competition between the stray field emanating from the domains, and the demagnetizing field within the DW\cite{Hagedorn1974}. For $h$=30~nm, very loose HBLs also appear around 45~mT. This is consistent with the estimated HBL width  $\pi \Lambda$=$\pi\sqrt{Q}$=27~nm. A 40~nm layer can thus accommodate with difficulty a full HBL. 

To summarize, this first analysis gives two fundamental results: the kinks in the v(H) curves occur at fields maximizing both the curvature and the twist of the DW, and twist and curvature are coupled variables.

\section{ANALYSIS}

 In order to get more insight into the respective roles of the DW flexion and magnetization twist on the kinks of the v(H) curves, the analytical model of Malozemoff and Slonczewski\cite{malo79} will be used. Let us  beforehand remind the general equation coupling the time derivatives of  ($q_{0}$,$\phi_{0}$)(t) in the 1D model:

 \begin{equation}
 \dot{q_{0}}(t)=\frac{\mu_{0} M}{2} \gamma \Delta  \sin 2\phi_{0}+\alpha \Delta \dot{\phi_{0}}(t) 
 	\label{eq:q0}
   \end{equation}

In the precessional regime, the propagation is achieved by a balance of the torque associated to the oscillating demagnetizing field (first term of the right hand side of Eq. (6)) created by the precession, and the dissipation of the energy brought by the field in the form of a damping contribution (second term). For $H>>H_{W}$, the demagnetization term  averages out to zero, and the damping term takes over and propels the DW forward with a velocity proportional to the applied field. In this simple 1D model, two different routes appear to increase v(H): an increased  \textit{precession rate} or an increased \textit{damping}. As mentioned above however, the simulations showed no substantial difference for $\dot{\phi}$ between the different thicknesses. Concerning the damping,  an alternative way for the system to dissipate energy would be  by direct coupling to spin-waves (SW) in the bulk of the material, or by a spin-wave like excitation of the magnetization in the vicinity of the DW\cite{Hagedorn1961,Schlomann1971}. Plotting the $m_{x,y,z}$(y)  components for different thicknesses does not evidence any particular wake, but only a slight deformation of the DW profile for $H \approx H_{res}$. Concerning bulk spin-waves, their lowest ($k$=0) frequencies\cite{Farle1998} lie much higher than the typical precession frequencies, thus forbidding any direct coupling between the DW and bulk SWs. It therefore seems necessary to go beyond the 1D model, and take into account the full thickness dependence of the magnetization configuration.

For a twisted domain wall taken infinite in the $x$ direction, the DW surface energy is expressed to first order in exchange, magnetostatic, and curvature terms as\cite{malo79}:

\begin{align}
     \sigma & = \sigma_{0}\left[1+\frac{1}{2}\left(\nabla q\right)^{2}\right]+2A\Delta \left(\nabla \psi\right)^{2} + \\
\nonumber   &   \mu_{0}M^2 \Delta \sin^{2}\psi- \Delta \mu_{0}\pi M H_{y}\sin \psi- 2 \mu_{0}M Hq
 	\label{eq:sigma}
\end{align}

In this expression, $\sigma_{0}=4\sqrt{AK}$ is the 1D  DW surface energy. The  term  $\left[1+\frac{1}{2}\left(\nabla q\right)^{2}\right]$ corresponds to the energy increase due to the elongation of the DW, and the term $A\left(\nabla \psi\right)^{2} $ corresponds to the exchange energy cost arising from the twist. By analogy with a spring or a twisted rope, and looking at Fig. \ref{fig:Fig3}b, the DW energy contains a kinetic (flexion $\nabla^{2}q$) and an elastic (twist $\nabla ^{2}\psi$) term which balance each other out. When the DW is very twisted, the energy cost of an additional curvature is too high: the flexion is minimal, and vice-versa. $H_{y}$ can stem from volume or surface magnetic charges created by the DW or domains. Here, we will only consider $H_{y}$ as the  $y$ component of the stray field arising from up- and down-domains on either side of the DW. The curvature of the domain will be considered small enough to neglect the fields along $y$ and $z$ resulting from  magnetic charges appearing along the DW due to its curvature\cite{malo79}. Among the different expressions demonstrated for $H_{y}$\cite{Hagedorn1974,Hubert1975}, we will use that of Ref.~\onlinecite{Hagedorn1974}, as it remains valid down to the $\Delta \approx h$ limit.

\begin{align}
 4\pi H_{y}(z)     & =  -2M\ln\left[\frac{z^{2}+\Delta^{2}/4}{(h-z)^{2}+\Delta^{2}/4}\right]+\\
    \nonumber &    \frac{8M}{\Delta}\left[z\arctan  \left(\frac{\Delta}{2z}\right)-(h-z)\arctan \left(\frac{\Delta}{2(h-z)}\right)\right]
    \label{Hy}
\end{align}

It is determined by computing the potential arising from the -$M$ and +$M$ surface charges (varying linearly  over a width $\Delta$) of domains on either side of the DW. Any effect of the volume charges created within the DW are ignored. This field is stationary, but varies throughout the thickness of the material. It pins the magnetization of a static DW in N\'{e}el configuration at the surfaces of the layer, so that the equilibrium structure of the wall is twisted. For $h$=30~nm, it amounts to a field of over 55~mT at the surface of the layer. This is likely an overestimation of the actual $H_{y}$ though, since the volume  contribution has been ignored. Finally, using the LLG equation leads to the  equation\cite{malo79}:
 
\begin{widetext} 
  
\begin{eqnarray}
\dot{q}(z,t)  & =& \frac{\gamma \Delta}{1+ \alpha^{2}} \left[\frac{\mu_{0} M}{2} \sin 2\psi  +\alpha \mu_{0} H 	-\frac{ \pi}{2}\mu_{0}H_{y}cos \psi-\frac{2 A \nabla ^{2}\psi}{M}+\frac{\alpha \sigma_{0} \nabla^{2}q}{2 M}\right]
 	\label{eq:qdot}
\end{eqnarray}

\end{widetext}

The DW velocity is then obtained by a double integration on time and space v(H)=$\frac{1}{T}\int^{T}_{t=0}\frac{1}{h}\int^{h}_{z=0}\dot{q}(z,t)dz\:dt$, provided the canting angle is known. As expressed in Eqs. (3)-(5), $\psi(z,t)$ can easily be written as the sum of the 1D model uniform $\phi_{0}$(t) given by Eq. \eqref{eq:psi0}, and a depth-dependent angle $\tilde{\psi}$(z,t) varying sinusoidally across the layer (in the absence of HBL). Moreover, the oscillation  amplitude $S_{0}(H)$ can be  fitted by a Lorentzian (Fig. \ref{fig:Fig2}c, solid line). The negative signs in Eqs. (4,5)  mimic the  phase seen in the simulations:  the twist is null when the wall is in a Bloch configuration ($\phi_{0}$=0 [180$^{\circ}$]), and extremum when it is in a pseudo Neel configuration $\phi_{0}$=90$^{\circ}$ [180$^{\circ}$]).

\begin{figure}
	\centering
				\includegraphics[width=0.50\textwidth]{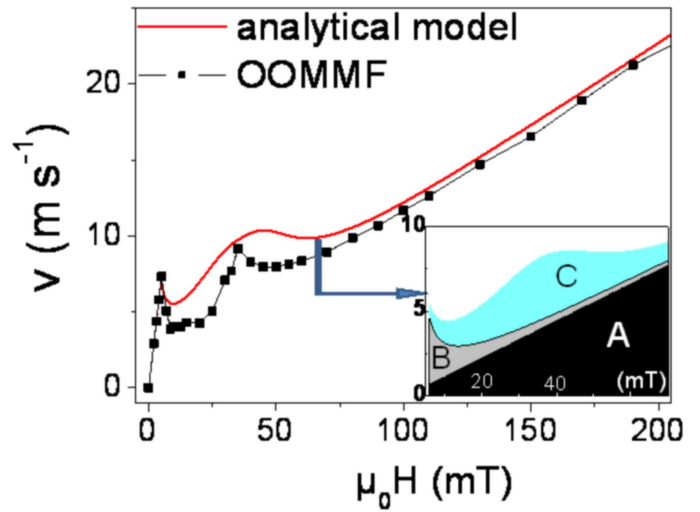}
	\caption{[color on-line] Velocity computed analytically using the ($q$, $\psi$, $\nabla q$, $\nabla \psi$) model for $h$=30~nm and $H>H_{W}$, and the corresponding simulated OOMMF curve. [inset] Close-up of the kink area evidencing the respective contributions to the velocity of the applied field ($A$), the demagnetizing field ($B$) and the stray field ($C$). See Eq. (9) and text for details.}
	\label{fig:Fig4}
\end{figure}

The resulting computed velocity is plotted in Fig. \ref{fig:Fig4} for $h$=30~nm, along with the corresponding micromagnetic simulation already presented in Fig. \ref{fig:Fig2}c.  The Walker field and DW thickness injected in Eq.~(2) are those of the 1D model baseline fit of the $h$=30~nm curve : $\mu_{0} H_{W}$=5~mT and $\Delta$=2.28~nm. The v(H) curve obtained using the analytical model given by Eqs.~(\ref{eq:sigma}-\ref{eq:qdot}) resembles very much the numerical simulation. The mobility above Walker breakdown is linear for the 1D model curve, but shows a kink when the magnetization twist is taken into account. At the peak ($\mu_{0} H\approx$38~mT), the velocity has increased from 4.7~m~s$^{-1}$ to about 10~m~s$^{-1}$, to be compared to the value of 9.2~m~s$^{-1}$ obtained in the simulation. At high fields, the calculated curve eventually coincides with the 1D model one. The kink is in fact broader that the one observed in the simulation, which is not surprising considering the approximations needed to establish Eq. (8). A very similar result is obtained for the $h$=20~nm curve, showing that the velocity increase is due to the twist of the magnetization.

A partial analytical integration of Eq.~(\ref{eq:qdot}) over $t$ and $z$  gives some insight as to which of the three terms $H_{y}$, $\nabla^{2} q$ and/or  $\nabla^{2} \psi$ - absent from the 1D model expression - governs the  shape of v(H). Taking into account the
boundary conditions  $\frac{\partial \psi}{\partial z}|_{0,h}$=$\frac{\partial q}{\partial z}|_{0,h}$=0  immediately shows that the $\nabla^{2} q$ and  $\nabla^{2} \psi$ terms do not contribute at all. Using the  odd-nature about $z=h/2$ of both $H_{y}$ and $\tilde{\psi}$ then leaves:

\begin{widetext} 
\begin{equation}
<v>  = \frac{\Delta \gamma }{1+ \alpha^{2}}\left\{ \frac{\mu_{0} M}{2} \frac{1}{T} \int^{T}_{0}J_{0}(2S_{0}\sin\frac{2\pi t}{T})\sin 2\phi_{0}dt +
     \frac{ \pi}{2\; T h} \int^{T}_{0}\sin \phi_{0}\int^{h}_{0}\mu_{0}H_{y}(z)\sin \tilde{\psi}dzdt +\alpha \;\mu_{0} H_z\right\}
\label{eq:vH}
\end{equation}
\end{widetext}

$J_{0}$(x) refers to the $m$=0 Bessel function of the first kind. In the absence of twist within the DW ($S_{0}$=0), the $\tilde{\psi}$ term vanishes, and the 1D expression is recovered. In the absence of time dependence of the twist ($S$(t)=$S_{0}$), the second term in Eq.(\ref{eq:vH}) vanishes by time integration of the T-periodic $\sin \phi_{0}$ term. This leaves an expression very similar to the 1D equation, save for a  decrease by $J_0(2S_{0})$ of the demagnetization term. 

When the twist is taken to be time dependent as described in Eqs.~(3)-(5), the demagnetizing field (first term of Eq. (10)) created across the domain wall over one period is decreased compared to the pure Bloch DW case, as previously. The second term of Eq. (10) reflects the strength of the torque $\vec{m}\times \vec{H}_{y}$ averaged over the thickness of the layer, and over a period.  Because both the stray field and the z-dependence of the DW magnetization $\tilde{\psi}$ invert signs at mid height of the layer, the integration over the layer thickness of this torque will always be non-zero. The sign of this  contribution is moreover given by the phase of the twist amplitude $S$(t) with respect to the mean azimuthal angle of the DW $\phi_{0}$(t). As shown in Fig. \ref{fig:Fig3}, when the twist is winding clockwise  up the layer ($S<0$), the magnetization lies opposite the direction of motion ($\phi_{0}$=0-180$^{\circ}$, sin$\phi_{0}>$0). When the twist is winding counterclockwise ($S>0$), the magnetization is facing  the direction of motion ($\phi_{0}$=180-360$^{\circ}$, sin$\phi_{0}<$0). As a result, this integration yields over a whole precession period a net  positive contribution to the DW velocity.  The  amplitude of $H_{y}$ can easily be of the order of 1.3$M$ at the surface, and so the $\vec{m}\times \vec{H}_{y}$ torque  will dominate the first term.  Plotting only the applied field and  demagnetizing field terms calculated with a twisted magnetization (labeled $A$ and $B$ in the inset Fig. \ref{fig:Fig4}) gives a curve very similar to the one computed from the 1D model equation (6). It is thus clearly the stray field term ($C$) that yields a velocity increase.

For an increased velocity to appear, it is therefore required to combine both a twist of the DW magnetization across the layer, and a particular time dependence of this twist amplitude. As shown above, the twist and curvature are coupled variables. At fields giving a maximum flexion, a maximum twist will therefore also be obtained. The position of the resulting kink in the v(H) curve will then be indirectly related to  the field dependence of the curvature $\eta(H)$, given by $\nabla^{2}q$.

Finally, we come to the double kink for $h$=40~nm. At low field the kink is very similar to the $h$=30~nm curve, but abruptly drops around $\mu_{0} H$=30~mT. Since the velocity increase is directly due to the continuous and symmetrical twist of the DW across the layer's width,  horizontal Bloch lines are likely to severely disrupt the scheme presented above, all the more so as the HBL is tighter. As shown in Fig. \ref{fig:Fig3bis}b, HBLs precisely start appearing around $\mu_{0}H$=30~mT for $h$=40~nm. They are then possibly responsible for splitting the peak into a sharp kink followed by a broad bump.

\section{POSITION OF VELOCITY PEAKS}

\subsection*{\label{sec:Simulations}A. Simulations\protect\\}

An important question remains concerning the  position of the velocity peaks, as well as their thickness dependence. Because the shape of $\eta_{0}(H)$ and $S_{0}(H)$ recalls a resonance, the free oscillation of a DW was investigated. With $H$=0 in the numerical simulation, the DW is launched from one of the configurations shown in the top caption of Fig. \ref{fig:Fig1}c: either a \textquotedblleft$n$=1\textquotedblright  sine mode, or a \textquotedblleft$n$=0" cosine mode, using once again the numbering of Slonczewski\cite{Slonczewski1981}. The DW is then let to relax. In both cases, its movement is initially highly non-linear, and then  rapidly falls into a damped oscillation of the $n$=1 mode. Higher orders of the DW are also occasionally observed. The frequency spectrum of the DW elongation obtained by fast Fourier transform is centered around the free oscillation frequency $f_{FO}$ of the DW. It is found to decrease when the thickness increases: $f_{FO}$=8.20, 2.34, 1.28 and 0.88~GHz  for respectively $h$=10, 20, 30 and 40~nm.

Let us now compare these frequencies with the precession frequency of the magnetization within the DW (solid line in Fig. \ref{fig:Fig5}, independent of layer thickness).  As shown by the arrows in Fig. \ref{fig:Fig5}, when the applied field induces a precession frequency close to $f_{FO}$, a velocity peak is obtained. The v(H) peak as well as the $\eta_{0}(H)$ therefore likely correspond to the resonance of the $n$=1 flexion mode of the DW. 

More exactly, the free oscillation resonance field falls slightly higher than the actual peak for $h$=20~nm (78~mT instead of $\mu_{0} H_{res}$=70~mT) and 30~nm (48~mT instead of $\mu_{0} H_{res}$=35~mT), and between the peak and the bump for $h$=40~nm (35~mT). An explanation as to why the peaks are observed at lower frequency than expected from the free oscillation simulations can be put forward. A domain wall in movement (under field) cannot be expected to have quite the same resonance frequency as a freely oscillating domain wall. In particular, applying an external field $H_{z}$ will disadvantage the higher order modes of the DW where magnetic charges due to its curvature will create an important magneto-static contribution. For $H_{z}=0$, the free oscillation frequency likely reflects the presence not only of the $n$=1 mode but also of higher order (and higher frequency) modes. These are indeed observed when plotting the DW depth-profile. These higher modes are progressively stifled upon increasing the field, and the frequency spectrum brought lower. This has been verified by doing  Fig. \ref{fig:Fig1}c-type simulations for $h$=20~nm under an applied field of 1, 3, 7.5 and 15~mT. As a result, the resonance under field appears at slightly lower frequency (and lower fields) than the free oscillation frequency. An interesting conclusion to this is that a DW cannot truly be considered as a harmonic oscillator as is often argued (Ref. \onlinecite{rhensius10} and therein), but rather as a parametric oscillator whose resonance frequency will effectively depend on the forcing function (applied field in our case).

\subsection*{B. Analytical model}
\label{sec:Simulations}

Slonczewski et al.\cite{Slonczewski1981} established an approximate expression for the flexural mode frequencies $f_{SL} $ of an assembly of domains arranged in stripes of infinite length (no applied field), neglecting damping. In our 2D geometry,  they can  be expressed as a function of $n$-indexed wave-vectors  $k_{z}(n)=n\pi/h$:

\begin{align}
f_{SL}  & = \frac{\gamma}{2\pi}\sqrt{\mu_{0} M+\frac{2 A}{M}k^{2}_{z}}\sqrt{\mu_{0}M\Delta k_{z}+\frac{2 A}{M}k^{2}_{z}}
     	\label{eq:fres}
\end{align}

\begin{figure}
	\centering
		\includegraphics[width=0.50\textwidth]{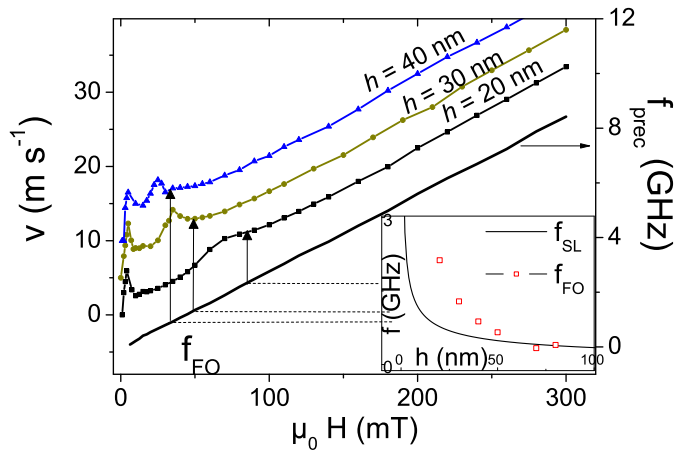}
	\caption{Velocity versus field curves (closed symbols, shifted vertically for clarity), and precession frequency (solid line): the fields giving the peaks in the v(H) curves coincide with the ones yielding the simulated free oscillation frequencies ($H$=0). [inset] Comparison between the thickness dependence of the  predicted resonance frequencies ($f_{SL}$) given in Ref. [\onlinecite{Slonczewski1981}] and the simulated ones ($f_{FO}$).}
	\label{fig:Fig5}
\end{figure}

The inset of Fig.~\ref{fig:Fig5} compares the simulated resonance frequencies determined above ($f_{FO}$, open squares), with  the frequency of the first $f_{SL}$ mode  computed from Eq. (11) with the numerical simulation's parameters. Since $f_{SL}$ was established at the first order in $\Delta k_{z}$, it largely underestimates the resonance frequencies for very thin layers, but both plots eventually converge for $h>70$~nm. Using Eq. (11) as a guideline   shows that for $h>>\Delta$ and $h\geq \Lambda$, the position of the peaks in v(H) (given by $f_{FO}$) will vary roughly  like $1/h$ with thickness and be very dependent on the $\Lambda/h$ ratio. It will not depend on the uniaxial anisotropy. This is to be compared to the $\sqrt{A}$ dependence of the  oscillation frequency of a transverse DW simulated numerically in Permalloy\cite{rhensius10}.  For sample thicknesses much larger than the exchange length, the frequency will  vary as $1/\sqrt{h}$, $A^{1/4}$ and $K^{-1/4}_{u}$. Although Eq. (11) can serve as a general guide line, one must however keep in mind that the restoring force in Ref. [\onlinecite{Slonczewski1981}] is somewhat different from  the one in our geometry.

\section{EXPERIMENTAL RESULTS}

This approach was then  confronted to  experimental data on perpendicularly magnetized samples. The dilute magnetic semiconductors GaMnAs and GaMnAsP offer weak pinning which allows the observation of both the stationary and the precessional DW propagation regimes\cite{Dourlat2008}. Moreover, magnetic parameters such as  $M$, $K_{u}$ and $A$, can  be adjusted by  doping\cite{Haghgoo2010,thevenard07,Cubukcu2010}. Contrary to the virtual material used in the simulations, GaMnAs also exhibits a complex \textit{in-plane} anisotropy, which can induce anisotropic velocities leading to spectacular non-circular domain shapes\cite{Gourdon2009}. However, the difference in velocity along the main axes being around 10~$\%$ only, the effect of the in-plane anisotropy is likely  much weaker than the mechanisms described above. 

\begin{figure}
	\centering
			\includegraphics[width=0.50\textwidth]{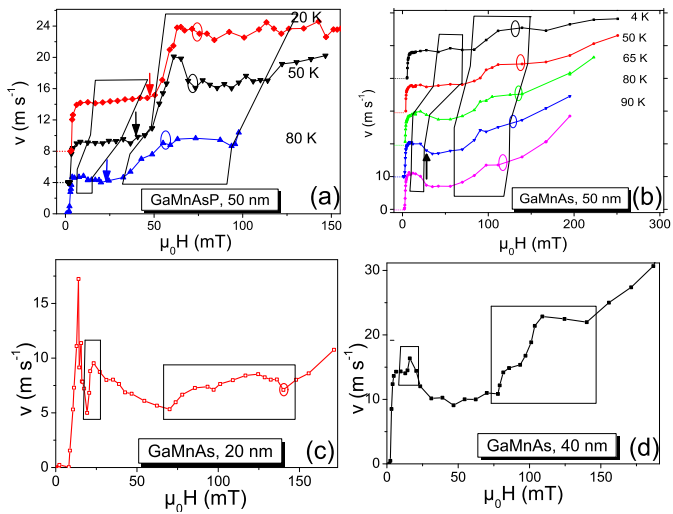}
	\caption{[color on-line] Experimental velocity versus field data (curves shifted for clarity where needed). The polygons indicate the velocity anomaly regions. The arrows point to the $H$=0 resonance fields determined numerically and the circles to the lower parametric excitation field (see text). (a) Ga$_{0.93}$Mn$_{0.08}$As$_{0.0915}$P$_{0.085}$,  $h$=50~nm. (b) Ga$_{0.93}$Mn$_{0.07}$As, $h$=50~nm (reproduced from Ref. [\onlinecite{Dourlat2008}]). (c) Ga$_{0.93}$Mn$_{0.07}$As  $h$=20~nm. ($T$=40~K), (d) and  $h$=40~nm ($T$=60~K).}
	\label{fig:Fig6}
\end{figure}

Fig. \ref{fig:Fig6}a shows the v(H) curve at 20, 50 and 80~K for a 50~nm thick Ga$_{0.93}$Mn$_{0.08}$As$_{0.915}$P$_{0.085}$ layer grown over GaAs. Fig. \ref{fig:Fig6}b,c,d show data for three Ga$_{0.93}$Mn$_{0.07}$As samples  grown on a In$_{y}$Ga$_{1-y}$As pseudo-substrate: $h$=50~nm at temperatures 4-90~K ($y$=15$\%$, previously published in Ref. [\onlinecite{Dourlat2008}]), and  $h$=20~nm ($y$=10$\%$, T=40~K) and $h$=40~nm ($y$=10$\%$, T=60~K). After annealing, these last two layers have Curie temperatures of respectively 118~K and 136~K. The micromagnetic parameters of both $h$=50~nm layers  were determined using Kerr microscopy as detailed in Ref. [\onlinecite{Haghgoo2010}] (where they are referred to as samples B and D). The main differences between these two samples are a higher uniaxial anisotropy in GaMnAs, and a higher exchange constant and saturation magnetization in GaMnAsP.  In all these measurements, the DW velocity was determined using magneto-optical imaging and a magnetic field pulse technique fully described in Ref. [\onlinecite{Dourlat2008}].

All the v(H) curves evidence low-field and high-field anomalies to the 1D model curve broadly indicated by polygons in Fig. \ref{fig:Fig6}. In the GaMnAs data, the $h$=50~nm curves show a weakly temperature-dependent kink at around 100-110~mT, and a discreet one at lower field (55~mT at $T$=4~K). The latter rapidly merges with the Walker peak ($\mu_{0} H_{W}\approx$5~mT) when temperature is increased. At high temperature, a  negative mobility  region is eventually obtained. The $h$=40~nm curve is  similar: a broad kink around Walker break-down ($\mu_{0} H_{W}\approx$5~mT), followed by a "split" bump centered around 100~mT. The $h$=20~nm curve exhibits distinctly Walker breakdown ($\mu_{0} H_{W}$=14~mT), a narrow negative mobility region before a peak at 23~mT. It is followed by a broad split bump centered around 105~mT.  The high field mobility is similar for $h$=40~nm and $h$=50~nm, and double that of the $h$=20~nm layer. To summarize, all GaMnAs samples exhibit a first velocity anomaly varying with temperature and thickness, and a second one independent of those parameters at $\approx$105~mT. The simulations ($h$=20~nm and 40~nm of Fig. \ref{fig:Fig2}b,d) reproduce qualitatively the low-field anomaly of the corresponding GaMnAs samples, but not the high-field 105~mT one. The GaMnAsP data (Fig. \ref{fig:Fig6}a) exhibits similarly two anomalous velocity features: a zero  or negative mobility regime  at low field and  a very broad bump  appearing at higher field: 50~mT at T=20, 50~K and 34~mT at T=80~K.

The resonance frequencies were then determined for each $h$=50~nm sample by micromagnetic simulations  ($\mu_{0}H$=0) as described above, using the experimental values of $M$, $K_{u}$ and $A$\cite{Haghgoo2010}. No sufficiently accurate value of $A$ was available for the $h$=20, 40~nm samples. The resonance fields found by comparing $f_{FO}$ to the precession frequency are indicated by arrows  directly  on the v(H) curves of Fig. \ref{fig:Fig6}a,b: 48, 40 and 23~mT for the GaMnAsP sample (T=20, 50 and 80~K), and 29~mT for the GaMnAs sample at $T$=80~K. Since the simulations showed that the low-field velocity anomalies appeared  slightly lower than the resonance fields, the arrows very likely correspond to the flexion/twist-induced kinks,  in good agreement with the description of the shifted peak position developed in Section IV. This would explain qualitatively the fact that the negative mobility regime appears upon increasing the temperature, as the flexion-induced kinks down-shift  towards the Walker field.  Taking into account the probable exchange constant inhomogeneity across the layer thickness would likely give better quantitative description of the kinks.

The $\approx$105~mT kinks in GaMnAs and $\approx$60~mT broad bumps  in GaMnAsP  can therefore  not be explained by  the twist/curvature model, as they shift only weakly  with  thickness or temperature. A tentative explanation for the origin of these kinks is the parametric excitation put forward by Randoshkin\cite{randoshkin86} whereby the bulk spin waves of the domains  couple  to the DW magnetization, thus providing an efficient energy dissipation channel leading to a velocity increase. This phenomenon is  expected  for frequencies equal to twice the precession frequency: $f_{bulk}$=2$f_{prec}$. This corresponds to resonance fields around $H_{u}$/3 or $2H_{u}$/3 depending whether the bulk spin waves are emitted in the domain oriented anti-parallel or parallel to the applied field respectively, and using the usual expression of bulk spin waves frequencies\cite{Herring1951}. When anisotropy coefficients were available, the lower bound expression could be evaluated (circles on the curves of Fig. \ref{fig:Fig6}a,b,c). The higher bound term falls out of the explored field range. For the GaMnAs samples, it yields about 110-140~mT ($T$=12-100~K) for $h$=50~nm,  $\approx$140~mT for  $h$=20~nm, and for  GaMnAsP 61-78~mT ($T$=20-80~K)  . These anharmonic coupling fields  follow  reasonably the temperature dependence of the  high-field bumps observed for both  GaMnAs and GaMnAsP, but do not explain the broadness of the bumps. The mechanism for this  coupling has moreover never been fully investigated. In the simulations, it would appear at around $H_{u}/3$=179~mT. It has not been observed, possibly due to the strong damping in the simulations.

\section{CONCLUSION}

To summarize, we have investigated numerically field-driven DW propagation in ferromagnetic layers of thickness close to the exchange length. These simulations reveal anomalous velocity peaks appearing in the precessional regime. They are correlated with the maximum flexion of the DW and the maximum twist of the magnetization vector inside the wall, both being intimately coupled. In order to elucidate their respective roles, we complement the historical ($q$, $\psi$) DW propagation model with the phenomenologically determined  $\nabla \psi$ (twist). The velocity peaks are then well reproduced and found to result  from the torque effect of the stray field on the twisted magnetization. When the DW magnetization is sufficiently twisted across the layer thickness, the stray field emanating from the up- and down-domains surrounding it induces an efficient torque on the total DW magnetization. Because the amplitude of this twist  varies over one precession period, the torque averages out to a strictly positive value. Anomalies in v(H) curves are therefore obtained for the particular range of thickness over exchange length ratio allowing a strong twist of the DW magnetization, but forbidding the appearance of horizontal Bloch lines, which tend to stifle the velocity enhancement.  The resonance field of the twist/curvature is such that the corresponding precession frequency is then close to the DW flexural resonance frequency, and  strongly depends on the layer thickness. The experimental v(H) curves obtained for ferromagnetic GaMnAs and GaMnAsP layers show velocity enhancement related to this DW magnetization twist and stray field-induced torque just above Walker break-down. At high field, they exhibit broad velocity bumps of different origin, which may be related to non-linear excitation of bulk spin-waves not evidenced by our 2D simulations. A full 3D simulation of the velocity would surely yield additional information.

The description of the velocity versus field curve in terms of the  ($q$,$\psi$,$\nabla q$,$\nabla \psi$) variables is an alternative and promising way to understand DW propagation experiments in perpendicular-to-plane magnetized samples. Keeping in mind that the  case of  Bloch DWs  is simpler in essence, the general ideas of this model also seem  applicable to in-plane magnetized materials. Velocity versus field curves taken in Permalloy nanostrips  have for instance very clearly evidenced kinks\cite{Moriya2010} in the precessional regime at fields decreasing with nanostrip \textit{width}, the relevant dimension for the flexion when $\vec{m}$ is in the plane of the sample. This would however call for more thorough investigation, as well as the relevance of the exposed model in current-induced DW propagation experiments\cite{Hayashi2006}. 

We acknowledge L. Largeau for the TEM determination of the GaMnAsP layer thickness. This work was in parts supported by R\'{e}gion Ile de France under contract IF07-800/R with C'Nano IdF.

\bibliographystyle{phjcp}
%\bibliography{library}

\begin{thebibliography}{10}
\bibitem{striplength}
{The strip length was long enough to avoid spurious effects arising from
  magnetic charges at the edges, except for fields 0-10mT for which a
  non-uniform field was applied in order to compensate this effect.}
\end{thebibliography}

\pagebreak[4]

\end{document}